\documentclass[12pt]{article}
\usepackage{amssymb}
\usepackage{amsfonts}
\usepackage{amsmath}
\usepackage[mathscr]{eucal}
\usepackage{amssymb}
\usepackage{amsthm}
\usepackage{bm}
\usepackage{graphicx}
\usepackage[english]{babel}
\usepackage[T2A]{fontenc}
\usepackage[utf8]{inputenc}
\usepackage{cite}

\def\theequation{\arabic{section}.\arabic{equation}}
\newcommand{\be}{\begin{equation}}
\newcommand{\ee}{\end{equation}}
\newcommand{\bea}{\begin{eqnarray}}
\newcommand{\eea}{\end{eqnarray}}

\newcommand{\lb}{\label}
\newcommand{\p}[1]{(\ref{#1})}

\newcounter{rown}

\begin{document}
\begin{titlepage}
\vspace*{0.1cm}

\begin{center}
{\LARGE\bf Continuous spin superparticle model}

\vspace{1cm}

{\large\bf I.L.\,Buchbinder$^{1,2,3}$\!\!,\ \ \
S.A.\,Fedoruk$^1$}

\vspace{1cm}

\ $^1${\it Bogoliubov Laboratory of Theoretical Physics,
Joint Institute for Nuclear Research, \\
141980 Dubna, Moscow Region, Russia}, \\
{\tt buchbinder@theor.jinr.ru, fedoruk@theor.jinr.ru}

\vskip 0.5cm

\ $^2${\it Center of Theoretical Physics,
Tomsk State Pedagogical University, \\
634041, Tomsk, Russia}, \\
{\tt joseph@tspu.edu.ru}

\vskip 0.5cm

\ $^3${\it National Research Tomsk State University,\\
\it 634050, Tomsk, Russia}\\

\end{center}

\vspace{2cm}

\nopagebreak

\begin{abstract}
\noindent We construct a new model of a particle propagating in
$4D$, ${\cal N}=1$ superspace that describes the dynamics of a
continuous spin irreducible representation of the Poincar\'{e}
supergroup. The model is characterized by two-component Weyl spinor
additional even variables playing the role of extra coordinates.
A canonical formulation, specific local fermionic $\kappa$-symmetry,
and a compete system of bosonic and fermionic constraints are derived.
All bosonic constrains are first-class, while fermionic
constraints are a mixture of first and second classes.
Using additional variables inherent in to the model, we
split the fermionic constraints into first and
second classes in a covariant way. Quantization of the model is carried out according to
Dirac prescription imposing all the first-class constraints and half of the second-class constraints
(Gupta-Bleuler procedure) on the wave function. At quantization, the
fermionic constraints are written in terms of spinor supercovariant
derivatives acting on superfields. The corresponding wave
function, which is either a chiral or antichiral superfield, depends on
additional variables and obeys the superfield constraints that
define the continuous spin irreducible representation of the
Poincar\'{e} supergroup in the superspace.
\end{abstract}

\vspace{2.5cm}


\phantom{Keywords: }

\newpage

\end{titlepage}

\setcounter{footnote}{0}
\setcounter{equation}{0}

\section{Introduction}

Relativistic particle models are
one-dimensional field theories that possess a number of global and
gauge symmetries. Such models can be viewed as an effective polygon
for studying numerous classical and quantum aspects of more
complicated field theories in different dimensions. At present,
there is an extensive  literature devoted to the formulation and
comprehensive study of particle models, containing at
least hundreds of titles. Among this literature, one can pick out a
class of massless superparticle models associated with particles
whose world lines lie in superspace (see the pioneering papers
\cite{Cas,BSch} and further developments
\cite{AzLuk-1982,Sieg-1983,Gupta-Bleuler,Gupta-Bleuler-1,Sieg-1988,BHT-1987}
as well as recent papers \cite{FIL-2006,BSa,KKR,KR} and references
therein).\footnote{Superparticle models should not be confused with
spinning particle models. The latter models are characterized by
supersymmetry on the world lines but not by target space supersymmetry
(see e.g. \cite{BDZVH,HPPT}).}
It was within the framework of the
superparticle model that a new type of fermionic gauge symmetry
(${\kappa}$-symmetry) was first discovered in \cite{AzLuk-1982,Sieg-1983}.
Note that the superparticle models played a certain role in the construction of
superstring theory to test the complicated string
concepts using simple enough models as an  example (see, e.g.,\cite{GSW,P} and references therein).

Superparticle models are examples of systems with
constraints in phase space. Canonical quantization of such systems
according to Dirac is carried out by imposing the operators of all the
first-class constraints and half of the second-class constraints on the wave function (see. e.g.,
\cite{Cas,FIL-2006,Gupta-Bleuler,Gupta-Bleuler-1,GT}).\footnote{This
quantization scheme is sometimes
called the Gupta-Bleuler quantization procedure.}
It was shown that
canonical quantization of superparticle models leads to equations of
motion for free superfields in flat or curved
superspace.\footnote{See e.g., derivation of the extended superfield
equations of motion from superparticle quantization in
\cite{AzLuk-1982,Gupta-Bleuler,Gupta-Bleuler-1,FIL-2006,BSa}.}

All the above models of superparticles in flat superspace
correspond to irreducible massless representations of the Poincaré
supergroup with certain superhelicities.
Relatively recently, a research direction has emerged in field
theory associated with the field Lagrangian description of
irreducible massless representations of the Poincaré group with
continuous (infinite spin) (see e.g. the review \cite{BeSk}
and references therein). The relativistic bosonic infinite
(continuous) spin particle models were constructed in flat and
curved spacetimes \cite{BFIR,BFI-19,BFIK-24-1}. In particular, it
was shown that such a consistent particle model exists only
 in either dS or AdS spaces. In this paper we propose a new
continuous spin relativistic particle model possessing
target space supersymmetry which can naturally be called a continuous
spin superparticle model.

The consideration of the continuous spin superparticle model is related to some extent
to the recent paper \cite{KRST}, where
a certain spinning particle model
was recently considered in the context of continuous spin field theory.
This model is characterized by the worldline $1D$ supersymmetry,
but not by the target space $4D$ supersymmetry.
In this paper we construct a model of a superparticle
that has target $4D$, $\mathcal{N}=1$ supersymmetry and describes the $4D$, $\mathcal{N}=1$  supermultiplet of infinite spin.

Since the consideration of infinite-spin representations of the Poincaré supergroup in \cite{BKRX-2002},
there has been a very limited number of papers on realizing such supersymmetric states.
Namely, in \cite{Zin-2017,BKSZ-2019,KhZin-2020} the Lagrangian formulations for
$4D$, $\mathcal{N}=1$ supermultiplets are presented in terms of the frame-like multispinor formalism and unfolded approach.
The equations defining the representations of $4D$, $\mathcal{N}=1$ supersymmetry on the superfields
with additional commuting vector variable were studied in \cite{BGK-2019}.
In addition, the papers \cite{N-2020} studied the supersymmetric generalization of the continuous spin gauge theory
proposed in \cite{ShT}.
A field realization of the $4D$, $\mathcal{N}=1$ supersymmetry representations of the continuous spin
was obtained in \cite{BFI-19} by studying twistor particles of infinite integer and half-integer spins.
This twistor formulation of the continuous spin particle is directly related
to its space-time formulation, in which an auxiliary field coordinate is a commuting Weyl spinor.
The BRST-triplet Lagrangian formulation for $4D$, $\mathcal{N}=1$ infinite spin superfields
with an additional spinor coordinate was constructed in \cite{BFIK-22}.

The paper is organized as follows. In Section\,2 we briefly review the
construction of the $4D$ bosonic continuous spin  particle model,
focusing  on the form of the Lagrangian and the structure of the
constraints. It is important to emphasize here that each formulation of such a model requires the use of
additional variables for which it is convenient to choose commuting
two-component Weyl spinors. Section\,3 is devoted to the generalization
of the above bosonic model to the continuous spin particle model
propagating in $4D$, $\mathcal{N}=1$ superspace. The local fermionic
$\kappa$-invariance is established, the canonical formulation is
developed, and a complete system of bosonic and fermionic constraints
is derived. It is noted that all the bosonic constraints are
first-class ones, while the fermionic spinor constraints are a mixture of
first-and second-class constraints.
A covariant split of fermionic constraints into first- and
second-classes is impossible without the use of additional even spinor variables.
In some superparticle models, such additional variables are
absent in the Lagrangians and have to be introduced "by
hand" as an extra assumption. A remarkable feature
of continuous particle models is that additional variables
themselves are part of the coordinates and must be mandatory
elements of the models. Therefore, they can be used without
any extra assumptions to split the fermionic constraints into
first- and second-class constraints. In Section\,4 the quantization of
the particle model under consideration is carried out. At
quantization, fermionic constraints are realized in terms of
covariant spinor derivatives acting on superfields. The
corresponding wave function is a chiral or antichiral $4D$, ${\cal N}=1$
superfield depending on additional two-component Weyl spinors
as additional variables and obey a certain system of constraints.
In Section\,5, it is shown that the above constraints define the
irreducible continuous spin representation of the  Poincar\'{e}
supergroup in superspace. Section\,6 summarizes the
results obtained. Appendix contains the details of deriving the
eigenvalues of the superfield fourth order Casimir operator.

\setcounter{equation}{0}

\section{Bosonic continuous spin particle}

Before we begin to consider the model of a continuous spin
superparticle, we will give, following the papers
\cite{BFIR,BFI-19,BFIK-24-1}, a brief overview of the corresponding
model of a pure-bosonic continuous spin particle.

The first order Lagrangian of a free continuous spin particle within
the formalism using the spinor auxiliary coordinates is written as
follows:\footnote{We use spinor notations as in \cite{WessBagger,Ideas}.}
\begin{equation}
\label{L-sp}
L_{b} \ = \ p_a \dot x^a \ +  \ \pi_{\alpha} \dot \xi^{\alpha}  \ +  \ \bar\pi_{\dot\alpha} \dot {\bar\xi}^{\dot\alpha}  \ -  \ \frac12\,e\, l_0 \  + \
\lambda\, l \  + \  \tilde\lambda\, \tilde l\  + \  iA u\,,
\end{equation}
where $e(\tau)$, $\lambda(\tau)$, $\tilde\lambda(\tau)$, $A(\tau)$ are the Lagrange multipliers for
the constraints
\begin{eqnarray}
l_0 &:=& p_a p^a \ \approx \ 0 \,, \label{const-sp}\\ [5pt]
l &:=& p_a \left(\xi\sigma^a \bar\xi \right)-\bm{\mu}  \ \approx \ 0  \,,  \label{const-sp-1}\\ [5pt]
\tilde l &:=& p_a \left(\bar\pi\tilde\sigma^a \pi \right) -\bm{\mu} \ \approx \ 0  \,,  \label{const-sp-2}\\ [5pt]
u &:=& N -\bar N \ \approx \ 0 \,,  \label{const-sp-3}
\end{eqnarray}
where
\begin{equation}
\label{N}
N:=\xi^{\alpha}\pi_{\alpha}\,,\qquad \bar N:=\bar\pi_{\dot\alpha} {\bar\xi}^{\dot\alpha}\,.
\end{equation}
The parameter $\bm{\mu}\in\mathbb{R}$ in \p{const-sp-1}, \p{const-sp-2} is a nonzero
constant, $x^a$, $p_a$, $a=0,1,2,3$ are the real coordinates and momenta of the particle,
$\xi^\alpha$,\, $\bar\xi^{\dot\alpha}=(\xi^\alpha)^*$, $\alpha=1,2$ are the spinor auxiliary coordinates,
$\pi_\alpha$,\, $\bar\pi_{\dot\alpha}=(\pi_\alpha)^*$ are the corresponding momenta and $\tau$ is an evolution parameter: $x^a=x^a(\tau)$,
$p_a=p_a(\tau)$, etc. We use the standard notation for
$\tau$-derivatives: $\dot x^a$, $\dot \xi^{\alpha}$,
$\dot {\bar\xi}^{\dot\alpha}$. Constraints \p{const-sp-1}, \p{const-sp-2} are imposed to get an infinite-spin irreducible representation of the Poincar\'{e} group. We stress that in the case when we use auxiliary spinor coordinates, we should introduce the constraint $u\approx 0$ \p{const-sp-3}
that commutes with all Casimir operators of the Poincar\'{e} group. This constraint provides an equal number of dotted and undotted spinor indices
in the dynamical invariants and hence guarantees the ability to switch to the description in terms of only vector indices (see details in \cite{BFIR}).

The Lagrangian \p{L-sp} can be rewritten in the form:
\begin{equation}
\label{L-sp-1}
L_{b} \ = \ p_a \left[\dot x^a+ \lambda  (\xi\sigma^a \bar\xi  )+ \tilde\lambda (\bar\pi\tilde\sigma^a \pi  ) \right]
\ +  \ \nabla\xi^{\alpha} \pi_{\alpha}  \   +  \ \bar\pi_{\dot\alpha} \nabla{\bar\xi}^{\dot\alpha} \ -  \ \frac12\,e \, p^2
\  - \  (\lambda+\tilde\lambda)\bm{\mu}\,,
\end{equation}
where
\begin{equation}
\label{nabla-xi}
\nabla\xi^{\alpha}:=\dot \xi^{\alpha} +iA\xi^{\alpha}\,,\qquad
\nabla{\bar\xi}^{\dot\alpha}:=\dot {\bar\xi}^{\dot\alpha}-iA{\bar\xi}^{\dot\alpha}\,.
\end{equation}
Eliminating the vector $p_a$ in the Lagrangian \p{L-sp-1} from its equation of motion, we obtain the equivalent second order Lagrangian
\begin{equation}
\label{L-sp-2}
L_{b} \ = \ \frac{1}{2e}\,V^a V_a
\ +  \ \nabla\xi^{\alpha} \pi_{\alpha}  \   +  \ \bar\pi_{\dot\alpha} \nabla{\bar\xi}^{\dot\alpha}
\  - \  (\lambda+\tilde\lambda)\bm{\mu}\,,
\end{equation}
where
\begin{equation}
V^a:=\dot x^a+ \lambda (\xi\sigma^a \bar\xi ) + \tilde\lambda (\bar\pi\tilde\sigma^a \pi ) .
\end{equation}
The Lagrangian (\ref{L-sp-2}) describes the dynamics of the bosonic infinite spin particle in $4D$ flat space.

\setcounter{equation}{0}

\section{Continuous spin superparticle in superspace: Lagrangian, Hamiltonian, and constraints}

To construct a superparticle model \p{L-sp-2} that describes the dynamics of an infinite spin particle in $4D$ flat target superspace,
we introduce the anticommuting  Weyl spinor coordinates $\theta^{\alpha}$, $\bar\theta_{\dot\alpha}$,
which together with the bosonic coordinates $x^{a}$ form the $4D$, ${\cal N}=1$ flat superspace with the coordinates $z^A=(x^a,\theta^{\alpha},\bar\theta_{\dot\alpha})$.
In adddition, the bosonic dynamical variabels $\xi^\alpha$, $\bar\xi^{\dot\alpha}$,
$\pi_\alpha$, $\bar\pi_{\dot\alpha}$, $\lambda$, $\tilde\lambda$
are introduced. These variables are an inherent ingredient of continuous spin particle models and
were previously used in the purely bosonic Lagrangian \p{L-sp-2}.
The set of variables $z^A$, $\xi^\alpha$, $\bar\xi^{\dot\alpha}$,
$\pi_\alpha$, $\bar\pi_{\dot\alpha}$, $\lambda$, $\tilde\lambda$ is considered as time-dependent particle supercoordinates
in the Lagrangian formalism.

Supersymmetry transformations of the $4D$, ${\cal N}=1$ flat superspace coordinates are written in the standard form
\begin{equation}\lb{trans-z}
\delta x^a = i\theta\sigma^a{\bar\varepsilon}-i\varepsilon\sigma^a{\bar\theta}\,,
\qquad  \delta\theta^{\alpha}=\varepsilon^{\alpha}\,,\qquad  \delta\bar\theta_{\dot\alpha}=\bar\varepsilon_{\dot\alpha}\,,
\end{equation}
where $\varepsilon^{\alpha}$, $\bar\varepsilon^{\dot\alpha}=(\varepsilon^{\alpha})^*$ are the
odd global parameters.
The bosonic auxiliary spinor and scalar variables are inert with respect to supertranslations:
\be\lb{trans-xi-pi}
\delta\xi^{\alpha}=0\, \quad\delta\bar\xi^{\dot\alpha}=0\,,\qquad
\delta\pi_{\alpha}=0\,, \quad \delta\bar\pi_{\dot\alpha}=0\,,\qquad
\delta\lambda=0\,, \quad \delta\tilde\lambda=0\,.
\ee

Supersymmetric-invariant quantities are the Cartan $\omega$-forms, which are obtained by shifting
\begin{equation}
\label{om}
\dot x^a \qquad \rightarrow \qquad \omega^a \ := \ \dot x^a  +i\theta\sigma^a\dot{\bar\theta}-i\dot\theta\sigma^a{\bar\theta}\,,
\end{equation}
of ``bosonic velocities'' $\dot x^a$.
By making replacement \p{om}  in \p{L-sp-2},
we obtain the target space supersymmetric-invariant Lagrangian
\begin{equation}
\label{L-super-1}
L \ = \ \frac{1}{2e}\,\mathscr{W}^{\,a} \mathscr{W}_{a}
\ +  \ \nabla\xi^{\alpha} \pi_{\alpha}  \   +  \ \bar\pi_{\dot\alpha} \nabla{\bar\xi}^{\dot\alpha}
\  - \  (\lambda+\tilde\lambda)\bm{\mu}\,,
\end{equation}
where
\begin{equation}
\label{Omega-1}
\mathscr{W}^{\,a} \ := \ \omega^a+ \lambda (\xi\sigma^a \bar\xi ) + \tilde\lambda (\bar\pi\tilde\sigma^a \pi ) \,,
\end{equation}
and $\nabla\xi^{\alpha}$, $\nabla{\bar\xi}^{\dot\alpha}$ are defined in \p{nabla-xi}.
Here, like in subsection 2.1,\, the quantities $\pi_\alpha$,\, $\bar\pi_{\dot\alpha}=(\pi_\alpha)^*$ play the role of the momenta conjugate to auxiliary spinor coordinates $\xi^{\alpha},\, \bar{\xi}^{\dot{\alpha}}$,
as we will show shortly after analyzing the constraints in the constructed system.

One can show that the action corresponding to Lagrangian (\ref{L-super-1})
\begin{equation}
\label{act-super-1}
\displaystyle S=\int d\tau L
\end{equation}
is invariant under the
local fermionic $\kappa$-invariance
\begin{equation}\lb{kappa}
\begin{array}{c}
\delta_\kappa x^a =
i\delta_\kappa\theta\sigma^a{\bar\theta}-i\theta\sigma^a\delta_\kappa\bar\theta\,, \\ [6pt]
\delta_\kappa\theta^{\alpha}=\mathscr{W}^{\,a}\bar\kappa_{\dot\beta}\tilde\sigma_a^{\dot\beta\alpha}\,,
\quad  \delta_\kappa\bar\theta^{\dot\alpha}=\mathscr{W}^{\,a}\tilde\sigma_a^{\dot\alpha\beta}\kappa_{\beta}\,, \\ [6pt]
\delta_\kappa e = 4ie\big( \dot\theta^\alpha\kappa_\alpha-\bar\kappa_{\dot\alpha}\dot{\bar\theta}^{\dot\alpha}\big), \\ [6pt]
\delta_\kappa\xi^{\alpha}=0\,, \quad \delta_\kappa\bar\xi^{\dot\alpha}=0\,,\qquad
\delta_\kappa\pi_{\alpha}=0\,, \quad \delta_\kappa\bar\pi_{\dot\alpha}=0\,,
\qquad
\delta_\kappa\lambda=0\,, \quad \delta_\kappa\tilde\lambda=0\,,
\end{array}
\end{equation}
where $\kappa_{\alpha}(\tau)$ and $\bar\kappa_{\dot\alpha}(\tau)=(\kappa_{\alpha}(\tau))^*$ are the local odd parameters.
With respect to these $\kappa$-transformations, the first term
$\displaystyle\frac{1}{2e}\,\mathscr{W}^{\,a} \mathscr{W}_{a}$ in the Lagrangian \p{L-super-1} and the rest
$\displaystyle \nabla\xi^{\alpha} \pi_{\alpha}  + \bar\pi_{\dot\alpha} \nabla{\bar\xi}^{\dot\alpha}
-  (\lambda+\tilde\lambda)\bm{\mu}$
are invariant separately. It is worth pointing out that the other terms in \p{L-super-1}
have the same form as in the bosonic Lagrangian \p{L-sp-2}.
Note that the Lagrangian \p{L-super-1} is invariant with respect to both global
\p{trans-z}, \p{trans-xi-pi} and local \p{kappa} fermionic transformations
and global and local bosonic transformations of the Lagrangian \p{L-sp-2}.

The action \p{act-super-1} yields the following momenta of all $1D$ fields:
\begin{eqnarray}
&\displaystyle p_a = \frac{\partial L}{\partial \dot x^a}= \frac{\mathscr{W}_a}{e}\,, & \lb{1}\\ [5pt]
&\displaystyle  p_{\alpha} = \frac{\partial L}{\partial \dot \theta^\alpha}=-ip_a(\sigma^a\bar\theta)_\alpha\,,
\quad   \bar p^{\,\dot\alpha}= \frac{\partial L}{\partial \dot {\bar\theta}_{\dot\alpha}}=
-ip_a(\tilde\sigma^a\theta)^{\dot\alpha}\,, &  \lb{2}\\ [5pt]
&\displaystyle  p_{\xi\,\alpha} = \frac{\partial L}{\partial \dot \xi^\alpha}=\pi_{\alpha}\,,
\quad   \bar p_{\xi\,\dot\alpha}= \frac{\partial L}{\partial \dot {\bar\xi}^{\dot\alpha}}= \bar\pi_{\dot\alpha}\,, \qquad
p_{\pi}{}^{\alpha}= \frac{\partial L_{0}}{\partial \dot \pi_\alpha}=0\,,
\quad   p_{\pi}{}^{\dot\alpha}= \frac{\partial L}{\partial \dot {\bar\pi}_{\dot\alpha}}=0\,,&  \lb{3} \\ [5pt]
&\displaystyle  p_{e}= \frac{\partial L}{\partial \dot e}=0\,,
\quad   p_{A}= \frac{\partial L}{\partial \dot A}= 0\,, \quad
p_{\lambda}= \frac{\partial L}{\partial \dot \lambda}=0\,,
\quad   p_{\tilde\lambda}= \frac{\partial L}{\partial \dot{\tilde\lambda}}=0\,.&   \lb{4}
\end{eqnarray}
Spinor momenta satisfy the reality conditions:
\be
\bar p_{\xi\,\dot\alpha}=(p_{\xi\,\alpha})^*\,, \qquad \bar p_{\pi}{}^{\dot\alpha}=(p_{\pi}{}^{\alpha})^*
\ee
and
\be\lb{cc-p-th}
\bar p_{\dot\alpha}=(p_{\alpha})^*\,,\qquad
\mbox{where} \quad \bar p_{\dot\alpha}=\epsilon_{\dot\alpha\dot\beta}\bar p^{\dot\beta}\,.
\ee
Then the resulting phase space is restricted by a complete set of  constraints and
some of the phase space coordinates can be excluded.
After such a disposal of auxiliary  coordinates, some of the remaining fields have a natural interpretation
of either certain momentum variables or the Lagrange multipliers.\footnote{For details of the Hamiltonization procedure for the systems with constraints in phase space, see, e.g., \cite{GT}.}
Let us proceed to this fairly simple procedure.

The canonical Hamiltonian defined as $H_c=\dot Z^{\cal M}P_{\cal M}- L$, where $Z^{\cal M}$ and $P_{\cal M}$ are all the coordinates and their momenta, has the form
\begin{equation}
\label{H-sp}
H_{c} \ = \ \frac12\,e\, l_0 \  - \
\lambda\, l \  - \  \tilde\lambda\, \tilde l\  - \  iA u\,,
\end{equation}
where $l_0$, $l$, $\tilde l$, $u$ are defined in  \p{const-sp}-\p{const-sp-3}.

Expressions \p{2}-\p{4} give us primary constraints.
As well known, the total Hamiltonian (at the first stage) is the sum of the Hamiltonian \p{H-sp}
and a linear combination of the above primary constraints with the corresponding Lagrange multipliers.

Preserving the primary constraints \p{4}
$p_{e}\approx 0$,
$p_{\lambda}\approx 0$,
$p_{\tilde\lambda}\approx 0$,
$p_{A}\approx 0$
produces constraints \p{const-sp}-\p{const-sp-3} $l_0\approx 0$, $l\approx 0$, $\tilde l\approx 0$, $u\approx 0$
as secondary constraints.
At the same time, constraints \p{4} say that the variables $e(\tau)$, $\lambda(\tau)$, $\tilde\lambda(\tau)$, $A(\tau)$ become Lagrange multipliers for these secondary constraints.

Expressions \p{3} define the constraints  $p_{\pi}{}^{\alpha}\approx 0$,
$\bar p_{\pi}{}^{\dot\alpha}\approx 0$, $p_{\xi\,\alpha}-\pi_{\alpha}\approx 0$,
$\bar p_{\xi\,\dot\alpha}- \bar\pi_{\dot\alpha}\approx 0$, which have a simple form and
are second class constraints.
A simple procedure for introducing the Dirac bracket with respect to these constraints leads
to the removal of the variables
$p_{\xi\,\alpha}$, $\bar p_{\xi\,\dot\alpha}$, $p_{\pi}{}^{\alpha}$, $p_{\bar\pi}{}^{\dot\alpha}$ from the phase space.
After this, the quantities $\pi_\alpha$, $\bar\pi_{\dot\alpha}$  become the momenta conjugate to the bosonic spinor  coordinates $\xi^{\alpha},\, \bar{\xi}^{\dot{\alpha}}$, which is exactly the same as in the purely bosonic case of Sect.\,2.
For the remaining variables, these Dirac brackets coincide with the Poisson brackets.
For this reason, below, when considering this system, the canonical brackets are called Poisson brackets.

As a result, after processing the primary constraints \p{3} and \p{4},
the phase space of the considered system is described by
the bosonic $x^a$, $p_a$; anticommuting $\theta^{\alpha}$, $\bar\theta_{\dot\alpha}$, $p_{\alpha}$, $\bar p^{\dot\alpha}$ and auxiliary
$\xi^\alpha$, $\bar\xi^{\dot\alpha}$, $\pi_\alpha$, $\bar\pi_{\dot\alpha}$ coordinates and momenta.
They satisfy the canonical commutational relations in terms of the Poisson brackets \footnote{
Note that in our accepted notation, although $\big\{\theta^\alpha, p_{\beta} \big\} ={}- \delta^\alpha_\beta$ holds,
we have $\big\{\bar\theta^{\dot\alpha}, \bar p_{\dot\beta} \big\} = \delta^{\dot\alpha}_{\dot\beta}$.
But at the same time, the reality condition for odd momenta is $\bar p_{\dot\beta}=(p_{\beta})^*$ (see \p{cc-p-th}).}
\begin{eqnarray}
\label{PB0}
\big\{ p_{z^{A}} , z^B  \big\}={}-\delta_A^B\, :  &&
\big\{ x^a, p_b \big\}=\delta^a_b\,,\quad
\big\{\theta^\alpha, p_{\beta} \big\}={}-{}\delta^\alpha_\beta\,,\quad
\big\{\bar\theta_{\dot\alpha}, \bar p^{\dot\beta} \big\}={}-{}\delta_{\dot\alpha}^{\dot\beta}\,,
\\ [7pt]
\label{DB-b-sp}
&&\big\{\xi^\alpha, \pi_\beta \big\}=\delta^\alpha_\beta\,,\quad
\big\{\bar\xi^{\dot\alpha}, \bar\pi_{\dot\beta} \big\}=\delta^{\dot\alpha}_{\dot\beta}\,.
\end{eqnarray}
This phase space  is limited by a set of remaining constraints,
which contains even constraints \eqref{const-sp}-\eqref{const-sp-3}
and the odd constraints
\begin{equation}\lb{D-constr}
D_{\,\alpha}:=ip_{\alpha}-(\sigma^a\bar\theta)_\alpha p_a\approx 0\,,
\qquad   \bar D_{\,\dot\alpha}:=  i\bar p_{\dot\alpha}+(\theta\sigma^a)_{\dot\alpha}p_a\approx 0\,,
\end{equation}
where $\bar D_{\,\dot\alpha}=-(D_{\,\alpha})^*$, which
are a direct consequence of expressions \p{2}.

The non-vanishing Poisson brackets of the constraints \eqref{const-sp}-\eqref{const-sp-3} have the form
\begin{equation}
\label{algebra-1}
\big\{ l, \tilde l \big\} \ ={} \ -K l_0\,,
\end{equation}
where
\be\lb{K}
K \ := \ N +\bar N
\ee
and $N$ and $\bar N$ are given by \p{N}.
The constraints \eqref{const-sp}-\eqref{const-sp-3} have the vanishing Poisson brackets with constraints \p{D-constr},
whereas the non-vanishing Poisson brackets of constraints  \p{D-constr} are
\begin{equation}
\label{algebra-2}
\big\{ D_{\,\alpha}, \bar D_{\,\dot\alpha} \big\} \ = {} \ -2i \sigma^a_{\alpha\dot\alpha}\,p_a\,.
\end{equation}

The constraints \eqref{const-sp}-\eqref{const-sp-3} and \p{D-constr} make up a complete system of constraints in the model.
In fact, the constraints of an infinite spin superparticle are the sum of the constraints of an infinite spin particle,
and the constraints for the Brink-Schwartz superparticle with
the common constraint  $p^2\approx 0$.
Besides, the sectors of odd and even spinors relate to each other via the bosonic vector coordinates $x^a$, $p_a$.

Using expressions \p{algebra-1}, \p{algebra-2}, we find
that all bosonic constraints \eqref{const-sp}-\eqref{const-sp-3} are the first class constraints,
whereas four fermionic constraints \p{D-constr} are the mixture of two first class constraints and two second class constraints.
The first class constraints are extracted from the constraints by convolving them with the matrix $\tilde p^{\dot\alpha\alpha}$.
That is, the constraints\footnote{We use the standard notation: $p_{\alpha\dot\alpha}=\sigma^a_{\alpha\dot\alpha}p_a$,
$\tilde p^{\dot\alpha\alpha}=\tilde\sigma_a^{\dot\alpha\alpha}p^a$.}
\begin{equation}\lb{1D-constr}
F^{\,\dot\alpha}:=\tilde p^{\dot\alpha\alpha}D_{\,\alpha}\approx 0\,,
\qquad   \bar F^{\,\alpha}:= \bar D_{\,\dot\alpha}\tilde p^{\dot\alpha\alpha}\approx 0
\end{equation}
are the first class ones which are the generators of $\kappa$-transformations \p{kappa}.
However, constraints \p{1D-constr} are not independent due to  the constraint \eqref{const-sp}.
Extracting independent first- and second-class constraints from the constraints \p{D-constr} is impossible
in a covariant way without using additional commuting spinors.

However, a remarkable feature of the model under consideration is that it already contains commuting spinor variables $\xi^{\alpha},\,\bar{\xi}^{\dot{\alpha}}$ by construction. This opens up the opportunity to construct independent fermionic first- and second-class constraints in a covariant way using these variables
(see, e.g., a similar consideration in \cite{FIL-2006} in a different context).
As a result, the odd constraints \p{D-constr} are split into two second class constraints and two first class constraints as follows. The fermionic second class constraints look like
\begin{equation}
\lb{sG-constr}
G:=\xi^{\alpha}D_{\,\alpha}\approx 0\,,
\qquad   \bar G:= \bar D_{\,\dot\alpha}\bar\xi^{\dot\alpha}\approx 0\,,
\qquad \bar G={}-(G)^*
\end{equation}
while the fermionic first class constraints are
\begin{equation}\lb{fF-constr}
F:=\bar\xi_{\dot\alpha}\tilde p^{\dot\alpha\alpha}D_{\,\alpha}\approx 0\,,
\qquad   \bar F:= \bar D_{\,\dot\alpha}\tilde p^{\dot\alpha\alpha}\xi_{\alpha}\approx 0\,.
\qquad \bar F={}-(F)^*
\end{equation}
The algebra of constraints \p{sG-constr} and \p{fF-constr} has the form
\begin{equation}
\label{algebra-2-1}
\big\{ G, \bar G \big\} = {} -2i\, l -2i\bm{\mu}\,,\qquad
\big\{ F, \bar F \big\} = {}  2i (\xi p\,\bar\xi)\,l_0\,.
\end{equation}
They commute (in weak sense) with all bosonic constraints \p{const-sp}-\p{const-sp-3}:
\begin{equation}
\label{algebra-2a}
\big\{ G, \tilde l \big\} = F\,,\quad
\big\{ \bar G, \tilde l \big\} = \bar F\,, \qquad
\big\{ F, \tilde l \big\} = \pi^{\alpha}D_{\alpha}\,l_0\,,\quad
\big\{ \bar F, \tilde l \big\} = \bar D_{\dot\alpha}\bar\pi^{\dot\alpha}\,l_0\,,
\end{equation}
\begin{equation}
\label{algebra-2b}
\big\{ G, u \big\} = G\,,\quad
\big\{ \bar G, u \big\} ={} -\bar G\,, \qquad
\big\{ F, u \big\} ={} -F\,,\quad
\big\{ \bar F, u \big\} = \bar F\,.
\end{equation}
It is clearly seen from \p{algebra-2-1} that the only second-class constraints are $G$ and $\bar{G}$.

Pay attention to the direct relation of the fermionic
spinor constraints \p{D-constr} with the fermionic first-class \p{fF-constr}
and second-class \p{sG-constr} constraints. It is not difficult to show
that the constraints \eqref{const-sp} and \eqref{const-sp-1},
\textit{i.\,e.} $p^2\approx 0$ and $\xi p\bar\xi\approx \bm{\mu}\neq
0$, lead to the identities
\be
\label{delta}
\delta_{\alpha}^{\beta}=\frac{1}{\bm{\mu}}\left[\xi_{\alpha}(\bar\xi\tilde
p)^{\beta} +(p\bar\xi)_{\alpha}\xi^{\beta}\right],\qquad
\delta_{\dot\alpha}^{\dot\beta}=\frac{1}{\bm{\mu}}\left[\bar\xi_{\dot\alpha}(\tilde
p\xi)^{\dot\beta} +(\xi p)_{\dot\alpha}\bar\xi^{\dot\beta}\right],
\ee
that hold on the surface of the constraints \eqref{const-sp} and
\eqref{const-sp-1}. Using these identities, we present the
constraints \p{D-constr} in the form
$$
D_{\alpha}\approx\frac{1}{\bm{\mu}}\,\left[(p\bar\xi)_{\alpha}G+\xi_{\alpha}F\right]\,,\qquad
\bar D_{\dot\alpha}\approx\frac{1}{\bm{\mu}}\,\left[(\xi p)_{\dot\alpha}\bar G+\bar\xi_{\dot\alpha}\bar F\right]\,.
$$
We see that the constraints $G$  and $F$ are equivalent to $D_{\alpha}$ and
the constraints $\bar G$  and $\bar F$ are equivalent to $\bar D_{\dot\alpha}$.\footnote{Similar
covariant extraction of independent first- and second-class constraints from the constraints (\ref{D-constr})
with help of commuting variables was used, e.g., in \cite{ZimFed,FIL-2006}.
}

The full Hamiltonian of the system contains all first-class constraints.
That is, it is the sum of the Hamiltonian \p{H-sp} that is a linear combination
of even first-class constraints \p{const-sp}-\p{const-sp-3}
plus a linear combination of odd first-class constraints \p{fF-constr} with odd Lagrangian multipliers.

\setcounter{equation}{0}

\section{Quantization of continuous spin superparticle}

The standard quantization of relativistic particle models is
based on the Gupta-Bleuler procedure (see, e.g.,
\cite{FIL-2006,Gupta-Bleuler,Gupta-Bleuler-1} and references
therein). According to this procedure, we have to impose on the wave
function all the first class constraints and half of the complex
second class constraints which (in the weak sense, modulo constraints)
commute. The remaining complex-conjugated second-class constraints
together with all the first-class constraints are imposed on the
conjugated wave function. After quantization we should get an
on-shell condition for the free space-time field. Further we consider a
realization of this procedure in the model under consideration.

In accordance with the Gupta-Bleuler quantization procedure,
we impose on the wave function $\Phi$ all the constraints \eqref{const-sp} and \eqref{const-sp-1},
\be\label{const-eqs}
l_0 \,\Phi = 0 \,, \qquad
l \,\Phi = 0  \,,  \qquad
\tilde l \,\Psi = 0  \,,  \qquad
u \,\Phi = 0 \,,
\ee
and one of two sets of the odd constraints.
The first (chiral) set consists of constraints $F\Phi = 0$, $\bar F\Phi = 0$, $\bar G\Phi = 0$ which are equivalent to
\be\label{chir-const-eqs}
\bar D_{\dot\alpha} \,\Phi = 0 \,, \qquad
(\bar\xi \tilde p D) \,\Phi = 0  \,.
\ee
Second (anti-chiral) set consists of constraints $F\bar\Phi = 0$, $\bar F\bar\Phi = 0$, $G\bar\Phi = 0$
on the wave function $\bar\Phi$ which are equivalent to
\be\label{achir-const-eqs}
D_{\alpha} \,\bar\Phi = 0 \,, \qquad
(\bar D \tilde p \xi) \,\bar\Phi = 0  \,.
\ee

Under quantization, classical momenta are realized by the operators:
\be\lb{mom-real}
p_a={}-{}i\partial_a\,,\quad \pi_{\alpha}=-i\partial/\partial\xi^{\alpha}\,,\quad \bar\pi_{\dot\alpha}=-i\partial/\partial\bar\xi^{\dot\alpha}\,,\quad
p_{\alpha}={}-{}i\partial/\partial\theta^{\alpha}\,,\quad
\bar p_{\dot\alpha}=i\partial/\partial\bar\theta^{\dot\alpha}\,.
\ee
In this case, the constraints \p{D-constr} take the form of supercovariant spinor derivatives of supersymmetric field theory (see, e.g., \cite{Ideas})
\begin{equation}\lb{cov-der-D}
D_{\,\alpha}=\frac{\partial}{\partial\,\theta^{\alpha}}+i(\sigma^a\bar\theta)_\alpha \partial_a \,,
\qquad   \bar D_{\,\dot\alpha}= -\frac{\partial}{\partial\,\bar\theta^{\dot\alpha}}-i(\theta\sigma^a)_{\dot\alpha}\partial_a \,.
\end{equation}

In the chiral case, the constraints \p{const-eqs} and \p{chir-const-eqs} take the form:
\be\lb{eq-fl-1}
\partial^a\partial_a\,\Phi \ = \ 0\,,
\ee
\be\lb{eq-fl-2}
{}-i\,(\xi\sigma^a\bar\xi)\,\partial_a\,\Phi \ = \ \bm{\mu}\,\Phi\,,
\ee
\be\lb{eq-fl-3}
i\,\Big(\frac{\partial}{\partial\bar\xi}\,\tilde\sigma^a\frac{\partial}{\partial \xi}\,\Big)\,\partial_a\,\Phi \ = \ \bm{\mu}\,\Phi\,,
\ee
\be\lb{eq-fl-4}
\Big(\xi\frac{\partial}{\partial \xi}-\bar\xi\frac{\partial}{\partial\bar\xi}\,\Big)\,\Phi \ = \ 0\,,
\ee
\be\label{chir-eq}
\bar D_{\dot\alpha} \,\Phi = 0 \,,
\ee
\be\label{chir-ed-D}
(\bar\xi \tilde\sigma^a D)\,\partial_a\,\Phi = 0  \,.
\ee

It is evident that the general solution to equation \p{chir-eq} is the chiral superfield depending on auxiliary variables $\xi^{\alpha},\, \bar{\xi}^{\dot{\alpha}}$:
\be\lb{wf-rep-chir}
\Phi \ = \ \Phi(x,\xi,\bar\xi,\theta,\bar\theta) \ = \ \Phi(x_{{}_L},\xi,\bar\xi,\theta)\,,
\ee
where $x^a_{{}_L}=x^a+i\theta\sigma^a\bar\theta$.

The chiral superfield $\Phi(x_{{}_L},\xi,\bar\xi,\theta)$ has the
standard component expansion
\be
\lb{wf-rep-chir-1}
\Phi(x_{{}_L},\xi,\bar\xi,\theta)\, = \varphi(x_{{}_L},\xi,\bar\xi)
+ \theta^{\alpha}\psi_{\alpha}(x_{{}_L},\xi,\bar\xi) + \theta^2
F(x_{{}_L},\xi,\bar\xi),
\ee
where $\varphi(x,\xi,\bar\xi),\,
F(x,\xi,\bar\xi)$ are the scalar fields and
$\psi_{\alpha}(x,\xi,\bar\xi)$ is the spinor field. It is clear that
relations \p{eq-fl-1}-\p{eq-fl-4} and \p{chir-ed-D} impose the
restrictions on component fields. The first of these relations is
the equation of free dynamics for each component, relations
\p{eq-fl-2}-\p{eq-fl-4} coincide with ones for free bosonic infinite
spin fields. Relation \p{chir-ed-D} is essentially new and is due
to supersymmetry.

Let us show that conditions \p{eq-fl-1}, \p{eq-fl-2} and \p{chir-ed-D} lead to the equation
$(D\sigma^a)_{\dot{\alpha}}\,\partial_a\,\Phi = 0$.
We consider the left hand side of above equation and use the identity \p{delta}. As a result, one gets
\be
\lb{chir-ed-D-1}
\begin{array}{rcl}
(D\sigma^a)_{\dot{\alpha}}\,\partial_a\,\Phi &=& \displaystyle
D^{\alpha}\delta_{\alpha}^{\beta}\partial_{\beta\dot{\alpha}}\Phi =
\frac{1}{\bm{\mu}}\Big[D^{\alpha}\xi_{\alpha}(\bar{\xi}\tilde{\partial})^{\beta} +
D^{\alpha}(\partial\bar{\xi})_{\alpha}\xi^{\beta}\Big]\,\partial_{\beta\dot{\alpha}}\Phi =
\\ [6pt]
&=& \displaystyle
\frac{1}{\bm{\mu}}\Big(D\xi\Big)\big(\bar{\xi}\tilde{\partial}\partial\big)_{\dot{\alpha}}\Phi
+
\frac{1}{\bm{\mu}}\big(\xi\partial\big)_{\dot{\alpha}}\big(D\partial\bar{\xi}\big)\Phi=0\,.
\end{array}
\ee
Here we have used the constraints $\tilde{\partial}\partial\Phi \sim \partial^2\Phi=0$ and $D\partial\bar{\xi}\Phi=0.$
As a result, the equation
\be
\lb{chir-ed-D-2}
(D\sigma^a)_{\dot{\alpha}}\partial_a\Phi =0
\ee
is holds.

The chiral superfield $\Phi(x,\xi, \bar\xi, \theta, \bar\theta)$ is
given by the standard component expansion
\be \lb{component}
\Phi =
\varphi(x,\xi,\bar\xi)\, +
\theta^{\alpha}\psi_{\alpha}(x,\xi,\bar\xi)\, + \theta^2
F(x,\xi,\bar\xi)+ \ldots,
\ee
where $\ldots$ means the $\bar\theta$ - dependent terms. Substituting expansion \p{component} into
equation \p{chir-ed-D-2} and putting $\theta= 0,\, \bar\theta=0,$
one gets the component equations
\be
\lb{component1}
\tilde\partial^{\dot{\alpha}\alpha}\psi_{\alpha}(x,\xi,\bar\xi)=0,\,
\qquad \tilde\partial^{\dot{\alpha}\alpha}F(x,\xi,\bar\xi)=0.
\ee
The last equation means that the field $F(x,\xi,\bar\xi)$ is space-time
independent and can be taken as an arbitrary constant.
Note that for the chiral field $\Phi$ equations \p{chir-ed-D-2} lead to the condition
\be
\lb{chir-ed-D-22}
D^\alpha D_\alpha\Phi=C\,,
\ee
where $C$ is a parameter independent of the space-time coordinates.\footnote{A similar equation arises when quantizing the Brink-Schwarz superparticle as well, see, e.g., \cite{FIL-2006}. Also, we point out that the model under consideration includes as a mandatory element
additional spinor variables. The corresponding field theory, which should obey the constraints
(\ref{eq-fl-1})-(\ref{chir-ed-D}), cannot be any conventional supersymmetric field model.
Relation (\ref{chir-ed-D-22}), where the right-hand side contains an arbitrary function which depends on additional variables, is completely consistent with the above constraints.}
It is this condition at $C=0$ that was used in \cite{BGK-2019} to describe continuous spin superfields
in which the additional even variable was a space-time four-vector.
We emphasize once more that equations \p{component1} are due to the
consideration of the superparticle model with a commuting spinor as an additional coordinate of the superfield.

Let us summarize the conclusions about the obtained spectrum of the superparticle model under consideration.

On the mass-shell, the dynamic component fields are fields $\varphi(x,\xi,\bar\xi)$ and
$\psi_{\alpha}(x,\xi,\bar\xi)$ that obey the equations
\be\lb{eq-varphi}
\partial^a\partial_a\,\varphi  =  0\,,
\quad
{}-i\,(\xi\sigma^a\bar\xi)\,\partial_a\,\varphi  =  \bm{\mu}\,\varphi\,,
\quad
i\,\Big(\frac{\partial}{\partial\bar\xi}\,\tilde\sigma^a\frac{\partial}{\partial \xi}\,\Big)\,\partial_a\,\varphi  =  \bm{\mu}\,\varphi\,,
\ee
\be\lb{eq-psi}
\tilde\sigma^{\dot\alpha\alpha}_a\partial^a\,\psi_\alpha  =  0\,,
\quad
{}-i\,(\xi\sigma^a\bar\xi)\,\partial_a\,\psi_\alpha  =  \bm{\mu}\,\psi_\alpha\,,
\quad
i\,\Big(\frac{\partial}{\partial\bar\xi}\,\tilde\sigma^a\frac{\partial}{\partial \xi}\,\Big)\,\partial_a\,\psi_\alpha  =  \bm{\mu}\,\psi_\alpha\,,
\ee
\be\lb{eq-varphi-psi}
\Big(\xi\frac{\partial}{\partial \xi}-\bar\xi\frac{\partial}{\partial\bar\xi}\,\Big)\,\varphi  =  0\,,
\quad
\Big(\xi\frac{\partial}{\partial \xi}-\bar\xi\frac{\partial}{\partial\bar\xi}\,\Big)\,\psi_\alpha  =  0
\ee
due to the constraint equations \p{eq-fl-1}-\p{eq-fl-4}.
As noted in \cite{BFI-19}, when studying twistor infinite spin fields,
it is precisely on these fields with these equations of motion that the mass-shell $\mathcal{N}=1$ supersymmetry can be realized.

Note that although equations \p{chir-ed-D-2}, which are the direct consequence of the quantization, differ from the superfield equations  obtained in \cite{BGK-2019}; they are on-shell equivalent since both describe the  same irreducible representation of continuous spin. The superfield description of the $4D$, ${\cal N}=1$ supersymmetric continuous spin irreducible representation was given in \cite{BGK-2019} under definite subsidiary conditions, and the corresponding Casimir operator was studied. In the present paper, we use other subsidiary conditions. In the next section, we will calculate the Casimir operator using our subsidiary conditions and show that egenvalues of the Casimir operator in our case and in the case of \cite{BGK-2019} coincide.\footnote{However, one can expect that the off-shell descriptions in both cases under consideration can be different. It would be useful to construct a superfield Lagrangian description corresponding to both these cases  and study a possible difference.}

\setcounter{equation}{0}

\section{Superfield continuous spin irreducible representation}

Let us show that when equations \p{eq-fl-1}-\p{chir-ed-D} are
satisfied, the superfield \p{wf-rep-chir} describes the irreducible
continuous spin representation of the ${\cal N}=1$ Poincar\'{e}
supergroup. It means that if the conditions
\p{eq-fl-1}-\p{chir-ed-D} are fulfilled, the corresponding
superfield Casimir operators have the correct
eigenvalues.

We begin with constructing the ${\cal N}=1$ superfield Casimir
operators $C_2$ and $C_4$ under the conditions
\p{eq-fl-1}-\p{chir-ed-D}. The operator $C_2$ has the standard form
$C_2=p^2$, where in the representation \p{mom-real},
$p_a={}-{}i\partial_a$. Taking into account \p{eq-fl-1}, we evidently
get the masslessness condition $C_2=0$
on the superfield $\Phi(x,\xi, \bar\xi, \theta, \bar\theta)$ in the spectrum of the considered superparticle.

The fourth order superfield
Casimir operator $C_4$ in the massless case is written in the form (see,
e.g., \cite{Ideas}):
\be
\lb{Cas4} C_4={}-\frac12\,
Z^{\alpha\dot\alpha}Z_{\alpha\dot\alpha}\,,
\ee
where
\begin{equation}
\label{Z}
Z_{\alpha\dot\alpha}=W_{\alpha\dot\alpha}-\frac18\,[Q_{\,\alpha}, \bar Q_{\,\dot\alpha}]
\end{equation}
is a generalization of the Pauli-Lubanski pseudovector
\begin{equation}
\label{W}
W_{\alpha\dot\alpha}=iM_{\alpha\beta}P^{\beta}_{\dot\alpha} - i\bar{M}_{\dot\alpha\dot\beta}P^{\dot\beta}_{\alpha}
\end{equation}
in the massless case, where $P_{\alpha\dot\beta}=p_{\alpha\dot\beta}$ and the spinor components $M_{\alpha\beta}$
and $\bar{M}_{\dot\alpha\dot\beta}$  of the angular momentum operator in the chiral $\xi,\,\bar{\xi}$-dependent
superfield \p{wf-rep-chir} representation  are
\begin{equation}\label{Msp}
M_{\alpha\beta} = i\xi_{(\alpha}\frac{\partial}{\partial\xi^{\beta)}}+i\theta_{(\alpha}\frac{\partial}{\partial\theta^{\beta)}}\,,
\qquad
\bar{M}_{\dot\alpha\dot\beta} =
i\bar{\xi}_{(\dot\alpha}\frac{\partial}{\partial{\bar\xi}^{\dot\beta)}}
+i\bar{\theta}_{(\dot\alpha}\frac{\partial}{\partial{\bar\theta}^{\dot\beta)}}\,.
\end{equation}
In expression \p{Z} for the four-vector $Z_{\alpha\dot\alpha},$ the quantities
\begin{equation}\lb{Q}
Q_{\,\alpha}=\frac{\partial}{\partial\,\theta^{\alpha}}-i(\sigma^a\bar\theta)_\alpha \partial_a \,,
\qquad   \bar Q_{\,\dot\alpha}= -\frac{\partial}{\partial\,\bar\theta^{\dot\alpha}}+i(\theta\sigma^a)_{\dot\alpha}\partial_a
\end{equation}
are the supercharges of $\mathcal{N}=1$ supersymmetry.

The operators $Q_{\alpha}$, $\bar{Q}_{\dot{\alpha}}$,
$M_{\alpha\beta}$,
$\bar{M}_{\dot{\alpha}\dot{\beta}}$, $W_{\alpha\dot{\alpha}}$,
$Z_{\alpha\dot{\alpha}}$ satisfy a large number of commutation
relations which allows us to derive the equation
\be \lb{mu}
C_4\Phi
= \bm{\mu}^2\,\Phi\,.
\ee
Thus, the fourth order  Casimir operator
under the conditions \p{eq-fl-1}-\p{chir-ed-D} takes the  eigenvalue
$\bm{\mu}^2$ corresponding to the continuous spin irreducible
representation. The details of deriving equation \p{mu} are
given in the Appendix.

\setcounter{equation}{0}

\section{Summary}
Let us summarize the results obtained. We have constructed the
continuous spin superparticle model. The model is formulated in
terms of the $4D$, ${\cal N}=1$ superspace coordinates
$z^{A}=(x^{a},\theta^{\alpha},{\bar{\theta}}_{\dot{\alpha}})$
supplemented by the commuting Weyl spinor coordinates
$\xi^{\alpha}$, ${\bar{\xi}}^{\dot{\alpha}}$. The coordinates $z^{A}$,
$\xi^{\alpha}$, ${\bar{\xi}}^{\dot{\alpha}}$ form the target space of the
particle under consideration. All these coordinates are the function
of the evolution parameter $\tau$.
Thus, the worldline of such a continuous spin model is a curve in superspace.

The target space supersymmetric Lagrangian of the model is given by
expression \p{L-super-1} and has local fermionic $\kappa$
invariance \p{kappa}. The canonical procedure based on the Lagrangian
\p{L-super-1} leads to the bosonic Hamiltonian $H_{c}$ \p{H-sp} and to the system of
bosonic first-class constraints $l_0$, $l$, $\tilde l$, $u$ defined in
\p{const-sp}-\p{const-sp-3}. Besides, the model also possesses the odd fermionic
constraints $D_{\alpha}$, $\bar{D}_{\dot{\alpha}}$ \p{D-constr}
which are a mixture of the first- and second-class constraints. As is
well known, the explicit splitting of such constraints into the first and
second classes requires the use of additional variables. Fortunately,
continuous spin particle models, unlike conventional superparticle models
\cite{BSch,AzLuk-1982,Sieg-1983,Gupta-Bleuler,Gupta-Bleuler-1,Sieg-1988,BHT-1987},
automatically contain the necessary additional Lorentz-spinor
variables $\xi^{\alpha}$, ${\bar{\xi}}^{\dot{\alpha}}$. Using these
variables, we have constructed the covariant split of fermionic
constraints. As a result, the second - and first - class fermionic
constraints were derived in the forms \p{sG-constr} and
\p{fF-constr}, respectively. Note that the constraints \p{fF-constr}
define an irreducible set of $\kappa$-symmetry generators.

We have carried out quantization of the model according to the Gupta-Bleuler procedure
imposing the operators of all the first-class constraints and half
of the second-class constraints on the wave function which becomes either a
chiral or antichiral ${\cal N}=1$ superfield which also depends on
additional variables $\xi^{\alpha}$, ${\bar{\xi}}^{\dot{\alpha}}$. In the chiral case, a complete system of
constraints for the superfield has been obtained in the form of
equations
\p{eq-fl-1}-\p{chir-ed-D}.
It was also proved that these equations define the continuous spin
irreducible representation of the Poincar\'{e} supergroup.

The results obtained open the possibility for further
interesting developments in
various directions:
\begin{itemize}
\item
Constructing new Lagrangian ${\cal N}=1$ superfield theory whose
equations of motion will be compatible with the constraints
\p{eq-fl-1}-\p{chir-ed-D}.
A preliminary step in this direction was made in \cite{BFIK-22},
where the BRST triplet field formulation for the $\mathcal{N}=1$ supersymmetric continuous spin theory was
proposed in superfield form.
\item
Constructing a continuous spin superparticle in curved
superspace and deriving constraints on the superfield that
generalize the constraints
\p{eq-fl-1}-\p{chir-ed-D}
in flat superspace. Deriving the corresponding matter
superfield theory in curved superspace.
\item
Constructing an ${\cal N}=2$ continuous spin superparticle and deriving the
corresponding ${\cal N}=2$ superfield theory. We expect that in this
case we should use the harmonic superspace methods \cite{GIOS}.
\item
Studying the possibility of constructing a continuous spin string theory which we assume should contain
an infinite tower of only massless states.
\end{itemize}
We plan to consider all these items in our forthcoming works.

\section*{Acknowledgments}
We are grateful to Evgeny Ivanov for useful critical comments.

%\appendix

\section*{Appendix.\ \ Eigenvalues of the superfield fourth order \\
\phantom{.....}\qquad\qquad  Casimir operator}

\renewcommand\theequation{A.\arabic{equation}} \setcounter{equation}0

In this Appendix we present the details of deriving the equation \p{mu} on the basis of expressions
\p{Cas4}-\p{Msp}
and subsidiary conditions \p{eq-fl-1}-\p{chir-ed-D}.

The direct calculations show that the operators \p{Q} and \p{Msp} satisfy the following relations in terms of anticommutators:
\begin{equation}\lb{Q-comm}
\{Q_{\,\alpha}, \bar Q_{\,\dot\alpha}\}={} -2 P_{\alpha\dot\alpha}\,, \qquad
[M_{\alpha\beta}, Q_{\,\gamma}]= i\epsilon_{\gamma(\alpha} Q_{\,\beta)}\,, \qquad
[\bar{M}_{\dot\alpha\dot\beta}, \bar Q_{\,\dot\gamma}]= i\epsilon_{\dot\gamma(\dot\alpha} \bar Q_{\,\dot\beta)} \,.
\end{equation}
The relations \p{Q-comm} and definition \p{W} allow one to derive
\begin{equation}\label{W-Q}
[W_{\alpha\dot\alpha},Q_{\,\beta}]={}-\epsilon_{\alpha\beta}(QP)_{\dot\alpha}-\frac12\,P_{\alpha\dot\alpha}Q_{\beta}\,,\qquad
[W_{\alpha\dot\alpha},\bar Q_{\,\dot\beta}]={}\epsilon_{\dot\alpha\dot\beta}(P\bar Q)_{\alpha}+\frac12\,P_{\alpha\dot\alpha}\bar Q_{\dot\beta}\,,
\end{equation}
and hence
\begin{equation}\label{W-Q-c}
[W_{\gamma\dot\alpha},Q^{\,\gamma}]={}\frac32\,(QP)_{\dot\alpha}\,,\qquad
[W_{\alpha\dot\gamma},\bar Q^{\,\dot\gamma}]={}-\frac32\,(P\bar Q)_{\alpha}\,.
\end{equation}
In its turn, the relations \p{W-Q} yield
\begin{equation}\label{Z-Q}
[Z_{\alpha\dot\alpha},Q_{\,\beta}]={}-\frac12\,\epsilon_{\alpha\beta}Q^{\,\gamma}P_{\gamma\dot\alpha}\,,\qquad
[Z_{\alpha\dot\alpha},\bar Q_{\,\dot\beta}]={}-\frac12\,\epsilon_{\dot\alpha\dot\beta}P_{\alpha\dot\gamma}\bar Q^{\,\dot\gamma}\,.
\end{equation}
Note that for unitary massless representations \cite{Ideas} the following relations
\be\lb{Q-ed1}
Q^{\,\gamma}P_{\gamma\dot\alpha}=0\,,\qquad P_{\alpha\dot\gamma}\bar Q^{\,\dot\gamma}=0\,
\ee
are fulfilled. Then, using
\be\lb{Q-D-ed}
Q_{\alpha}=D_{\alpha}+2(p\bar\theta)_{\alpha}\,,\qquad
\bar Q_{\dot\alpha}=\bar D_{\dot\alpha}-2(\theta p)_{\dot\alpha}\,,
\ee
we see that conditions \p{Q-ed1} are first-class constraints \p{1D-constr} on the massless shell $p^2=0$.

As a result of \p{W-Q-c} and \p{Q-ed1}, the vector $Z_{\alpha\dot\alpha}$ commutes with the supercharges.
Besides, $Z_{\alpha\dot\alpha}$ evidently commutes with the translation generators $[Z_{\alpha\dot\alpha},p_{\beta\dot\beta}]=0$.
The operator $Z^2:=Z^a Z_a$ is a scalar and as any scalar commutes with the  Lorentz rotation generators $M_{\alpha\beta}$, $\bar{M}_{\dot\alpha\dot\beta}$.
Thus, the operator
\be\lb{Cas-4}
C_4=Z^2={}-\frac12\, Z^{\alpha\dot\alpha}Z_{\alpha\dot\alpha}
\ee
is actually the fourth order Casimir operator.

Now let us  turn to calculating the eigenvalues of the Casimir operator $C_4$
on the chiral superfields $\Phi(x,\xi,\bar{\xi},\theta, \bar{\theta})$
\p{wf-rep-chir} under the subsidiary conditions \p{eq-fl-1}-\p{chir-ed-D}.

First of all, one writes the operator $Z^2$ in explicit form
\begin{eqnarray}
Z^{\alpha\dot\alpha}Z_{\alpha\dot\alpha}&=&-2 p^{\dot\alpha\alpha}p^{\dot\beta\beta}M_{\alpha\beta}\bar M_{\dot\alpha\dot\beta}
+\left(M^{\alpha\beta}M_{\alpha\beta}+\bar M_{\dot\alpha\dot\beta}\bar M^{\dot\alpha\dot\beta}-\frac18\right)p^2
\lb{Z2-calc}\\ [5pt]
&&{}+\frac14\,(Qp\bar Q) +\frac{1}{16}\,Q^{\alpha}Q_{\alpha}\bar Q_{\dot\alpha}\bar Q^{\dot\alpha}
+\frac{i}2\, M_{\alpha}^{\beta}Q^{\alpha}(p\bar Q)_{\beta}+
\frac{i}2\, \bar M_{\dot\alpha}^{\dot\beta}\bar Q^{\dot\alpha}(Qp)_{\dot\beta}\,. \nonumber
\end{eqnarray}
Substituting explicit expressions for the generators \p{Msp}, \p{Q} and relations \p{Q-D-ed} into \p{Z2-calc}
and taking into account that $p^2=0$, one obtains,
\begin{eqnarray}
Z^{\alpha\dot\alpha}Z_{\alpha\dot\alpha}&=&{}2(\xi p\bar\xi)\big(\frac{\partial}{\partial\bar\xi}\tilde p\frac{\partial}{\partial\xi}\big) +
2(\theta p\bar\theta)(\bar D\tilde p D) +
2(\xi p\bar\theta)\big(\bar D\tilde p\frac{\partial}{\partial\xi}\big) +
2(\theta p\bar\xi)\big(\frac{\partial}{\partial\bar\xi}\tilde p D\big) \nonumber \\ [5pt]
&&{}+\frac14\,(DP\bar D)
+\frac14\,(D^{\alpha}D_{\alpha}-4\bar\theta\tilde p D)(\bar D_{\dot\alpha}\bar D^{\dot\alpha}-4\theta\tilde p\bar D)  \nonumber \\ [5pt]
&&{}+\frac{i}2\, M_{\alpha}^{\beta}\big(D^{\alpha}-2(\theta p )^{\alpha}\big)(p\bar D)_{\beta}+
\frac{i}2\, \bar M_{\dot\alpha}^{\dot\beta}
\big(\bar D^{\dot\alpha}+2(\tilde p\theta)^{\dot\alpha}\big)(D p)_{\dot\beta}\,.
\lb{Z2-2}
\end{eqnarray}
Using relation \p{delta}
that holds on the surface of the constraint \p{eq-fl-2},
we find that the constraints \p{chir-eq}, \p{chir-ed-D} lead to the vanishing of all terms
on the right-hand side of \p{Z2-2}, with the exception of the first term,
when acting on the superfield \p{wf-rep-chir}.
That is, we have
\be
Z^2\,\Phi =
-(\xi p\bar{\xi})\left(\frac{\partial}{\partial \bar\xi^{\dot\beta}} p^{\dot\beta\beta}\frac{\partial}{\partial \xi^{\beta}}\right)\Phi\,.
\ee
But now, using \p{eq-fl-2} and \p{eq-fl-3}, one obtains
\be
Z^2\,\Phi =
\bm{\mu}^2\,\Phi\,.
\ee
As a result we see that the fourth-order Casimir operator takes the eigenvalue $\bm{\mu}^2$
corresponding to the irreducible continuous spin representation.


\begin{thebibliography}{96}

\bibitem{Cas}
R.\,Casalbuoni,
\textit{The classical mechanics for Bose-Fermi systems},
Nuovo Cim. A \textbf{33} (1976) 389

\bibitem{BSch}
L.\,Brink, J.H.\,Schwarz, \textit{Quantum Superspace}, Phys. Lett. B \textbf{100} (1981) 310.

\bibitem{AzLuk-1982}
J.A. de Azcarraga, J.\,Lukierski, \textit{Supersymmetric Particles with Internal Symmetries and Central Charges},
Phys. Lett. B \textbf{113} (1982) 170.

\bibitem{Sieg-1983}
W.\,Siegel,
\textit{Hidden Local Supersymmetry in the Supersymmetric Particle Action},
Phys. Lett. B \textbf{128} (1983) 397.

\bibitem{Gupta-Bleuler}
A.\,Frydryszak, \textit{N-extended free superfields (N=2,4,6,8) from
quantization of a supersymmetric particle model}, Phys. Rev. D
\textbf{30} (1984) 2172.

\bibitem{Gupta-Bleuler-1}
J.A.\,de\,Azcarraga, J.\,Lukierski, \textit{Supersymmetric particle
model with additional bosonic coordinates}, Z. Phys. C \textbf{30}
(1986) 221.

\bibitem{Sieg-1988}
W.\,Siegel, \textit{The superparticle revised}, Phys. Lett. B \textbf{203} (1988) 79.


\bibitem{BHT-1987}
L.\,Brink, M.\,Henneaux, C.\,Teitelboim,
\textit{Covariant Hamiltonian formulation of the superparticle},
Nucl. Phys. B \textbf{293} (1987) 505.

\bibitem{FIL-2006}
S.\,Fedoruk, E.\,Ivanov, J.\,Lukierski,
\textit{Massless higher spin D=4 superparticle with both N=1 supersymmetry and its bosonic counterpart},
Phys. Lett. B \textbf{641} (2006) 226, {\tt arXiv:hep-th/0606053}.

\bibitem{BSa}
I.L.\,Buchbinder, I.B.\,Samsonov,
\textit{${\cal N}=3$ superparticle model}, Nuc. Phys. B \textbf{802} (2008) 108, {\tt arXiv:0801.4907\,[hep-th]}.


\bibitem{KKR}
N.E.\,Koning, S.M.\,Kuzenko, E.S.N.\,Raptakis,
\textit{The anti-de Sitter supergeometry revised}, JHEP \textbf{02} (2025) 175, {\tt arXiv:2412.03172\,[hep-th]}.

\bibitem{KR}
N.E.\,Koning, E.S.N.\,Raptakis, \textit{New superparticle models in AdS superspaces}, {\tt arXiv:2506.17897\,[hep-th]}.

\bibitem{BDZVH}
L.\,Brink, S.\,Deser, B.\,Zumino, P.\,Di\,Vecchia, P.S.\,Howe,
\textit{Local supersymmetry for spinning particles},
Phys. Lett. B \textbf{64} (1976) 435.


\bibitem{HPPT}
P.\,Howe, S.\,Penati, M.\,Pernici, P.\,Townsend, \textit{Wave functions for srbitrary spin from quantization of the extended supersymmetric spinning particle,}
Phys. Lett. B \textbf{215} (1988) 555.


\bibitem{KRSchTor}
S.\,Kundu, A.\,Russo, P.\,Schuster, N.\,Toro,
\textit{Interactions of a continuous-spin field with a spin-1/2 particle},
{\tt arXiv:2505.14770\,[hep-th]}.

\bibitem{GSW}
M.\,Green, J.\,Schwarz and E.,Witten, \textit{Superstring Theory}, Vols I and II, Cambridge
University Press, 1987.

\bibitem{P}
J. Polchinski, \textit{String theory}, Vols I and II, Cambridge University Press, 2001.

\bibitem{GT}
D.M.\,Gitman, I.V.\,Tyutin, \textit{Quantization of Fields with Constraints}, Springer, 1990.

\bibitem{BeSk}
X.\,Bekaert, E.D.\,Skvortsov, \textit{Elementary particles with continuous spin}, Int. J. Mof. Phys. A \textbf{32} (2017) 1730019,
{\tt arXiv:1708.01030\,[hep-th]}.


\bibitem{BFIR}
I.L.\,Buchbinder, S.\,Fedoruk, A.P.\,Isaev, A.\,Rusnak,
\textit{ Model of massless relativistic particle with continuous spin and its twistorial description},
JHEP \textbf{07} (2018) 031, {\tt arXiv:1805.09706\,[hep-th]}.

\bibitem{BFI-19}
I.L.\,Buchbinder, S.\,Fedoruk, A.P.\,Isaev,
\textit{ Twistorial and space-time descriptions of massless infinite spin (super)particles and fields},
Nucl. Phys. B \textbf{945} (2019) 114660, {\tt arXiv:1903.07947\,[hep-th]}.

\bibitem{BFIK-24-1}
I.L.\,Buchbinder, S.A.\,Fedoruk, A.P.\,Isaev, V.A.\,Krykhtin,
\textit{Infinite (continuous) spin particle in constant curvature space},
Phys. Lett. B \textbf{853} (2024) 138689, {\tt arXiv:2402.13879\,[hep-th]};
\textit{BRST construction for infinite spin field on $AdS_4$},
Eur. Phys. J. Plus \textbf{139} (2024) 621,
{\tt arXiv:2403.14446\,[hep-th]}.

\bibitem{KRST}
S.\,Kundu, A.\,Russo, P.\,Schuster, N.\,Toro,
\textit{Interactions of a continuous-spin field with spin-1/2 particle}, {\tt arXiv:2505.14770\,[hep-th]}.


\bibitem{BKRX-2002}
L. Brink, A.M. Khan, P. Ramond, X.-z. Xiong,
\textit{Continuous spin representations of the Poincare and superPoincare groups},
J. Math. Phys. \textbf{43} (2002) 6279,
{\tt arXiv:hep-th/0205145\,[hep-th]}.

\bibitem{Zin-2017}
Yu.M.\,Zinoviev,
\textit{Infinite spin fields in d = 3 and beyond},
Universe \textbf{3} (2017) 63,
{\tt arXiv:1707.08832\,[hep-th]}.

\bibitem{BKSZ-2019}
I.L.\,Buchbinder, M.V.\,Khabarov, T.V.\,Snegirev, Yu.M.\,Zinoviev,
\textit{Lagrangian formulation for the infinite spin $N=1$ supermultiplets in $d=4$}
Nucl. Phys. B \textbf{946} (2019) 114717,
{\tt arXiv:1904.05580\,[hep-th]}.

\bibitem{KhZin-2020}
M.V.\,Khabarov, Yu.M.\,Zinoviev,
\textit{Massive higher spin supermultiplets unfolded},
Nucl. Phys. B \textbf{953} (2020) 114959, {\tt arXiv:2001.07903\,[hep-th]}.

\bibitem{BGK-2019}
I.L.\,Buchbinder, S.J.\,Gates, K.\,Koutrolikos,
\textit{Superfield continuous spin equations of motion},
Phys. Lett. B \textbf{793} (2019) 445,
{\tt arXiv:1903.08631\,[hep-th]}.

\bibitem{N-2020}
M.\,Najafizadeh,
\textit{Supersymmetric Continuous Spin Gauge Theory},
JHEP \textbf{03} (2020) 027,
{\tt arXiv:1912.12310\,[hep-th]};
\textit{Off-shell supersymmetric continuous spin gauge theory},
JHEP \textbf{02} (2022) 038,
{\tt arXiv:2112.10178\,[hep-th]}.

\bibitem{ShT}
P.\,Schuster, N.\,Toro,
\textit{A gauge field theory of continuous-spin particles},
JHEP \textbf{1310} (2013) 061, {\tt arXiv:1302.3225\,[hep-th]}

\bibitem{BFIK-22}
I.L.\,Buchbinder, S.A.\,Fedoruk, A.P.\,Isaev, V.A.\,Krykhtin,
\textit{On the off-shell superfield Lagrangian formulation of 4D, N=1 supersymmetric infinite spin theory}, Phys. Lett. B \textbf{829} (2022) 137139, {\tt arXiv:2203.12904\,[hep-th]};


\bibitem{WessBagger}
J.\,Wess, J.\,Bagger,
\textit{Supersymmetry and Supergravity}, Princeton Univ. Press, 1983.

\bibitem{Ideas}
I.L.\,Buchbinder, S.M.\,Kuzenko, \textit{Ideas and Methods of
Supersymmetry and Supergravity, Or a Walk Through Superspace}, IOP,
Bristol and Philadelphia, 1995 (Revised Edition 1998).

\bibitem{ZimFed}
V.G.\,Zima, S.\,Fedoruk
\textit{Spinor (super)particle with a commuting index spinor},
JETP Lett. \textbf{61} (1995) 251.

\bibitem{GIOS}
A.S.\,Galperin, E.A.\,Ivanov, V.I.\,Ogievetsky, E.S.\,Sokatchev,
\textit{Harmonic Superspace}, Cambridge Univ. Press, 2001.


\end{thebibliography}
\end{document}